\documentclass{llncs}
\usepackage{adjustbox}
\usepackage{graphicx,epsfig,color}
\usepackage[shortend,ruled]{algorithm2e}
\usepackage{ifthen,verbatim,array,multirow}
\usepackage{latexsym,amssymb,amsmath,amsthm}
\usepackage[english]{babel}
\usepackage[utf8]{inputenc}

\usepackage{xspace}
\usepackage{tikz}
\usetikzlibrary{arrows,decorations,positioning,shapes}
\usepackage{setspace}
\usepackage{enumitem}
\setlist{nolistsep}
\usepackage{listings}

\usepackage{textcomp}
\usepackage{stmaryrd}
\usepackage{pifont}
\usepackage{soul}

\usepackage{booktabs}   
\usepackage{multirow}   
\usepackage{cite}       
\usepackage{subcaption} 
\usepackage{todonotes}  
\usepackage{tablefootnote}
\usepackage[most]{tcolorbox}

\let\llncssubparagraph\subparagraph
\let\subparagraph\paragraph
\usepackage{titlesec}
\let\subparagraph\llncssubparagraph

\usepackage[pdftex%
,colorlinks=true       
,linkcolor=midblue        
,citecolor=midblue        
,filecolor=black       
,urlcolor=midblue         
,linktoc=page           
,plainpages=false]{hyperref}
\addto\extrasenglish{%

}

\fontfamily{ptm}

\setlist{nolistsep}
\setitemize{noitemsep,topsep=3pt,parsep=2pt,partopsep=2pt}
\setenumerate{noitemsep,topsep=3pt,parsep=2pt,partopsep=2pt}

\titlespacing{\section}{0pt}{*2.5}{*1.0}
\titlespacing{\subsection}{0pt}{*1.25}{*0.75}
\titlespacing{\subsubsection}{0pt}{*1.0}{*0.5}

\makeatletter
\renewcommand\paragraph{\@startsection{paragraph}{4}{\z@}%
                                    {1.0ex \@plus0.35ex \@minus.15ex}%
                                    {-1em}%
                                    {\normalfont\normalsize\bfseries}}

\makeatother

\makeatletter
\def\thm@space@setup{%
  \thm@preskip=1.5pt \thm@postskip=1.5pt
}
\makeatother

\makeatletter
\renewenvironment{proof}[1][\proofname]{\par
  \vspace{-0.25\topsep}
  \pushQED{\qed}%
  \normalfont
  \topsep0pt \partopsep0pt 
  \trivlist
  \item[\hskip\labelsep
        \itshape
    #1\@addpunct{.}]\ignorespaces
}{%
  \popQED\endtrivlist\@endpefalse
  \addvspace{1pt plus 1pt} 
}
\makeatother

\titleformat{\paragraph}[runin]
            {\normalfont\normalsize\bfseries}{\theparagraph}{0.4em}{}

\newtheorem{ntheorem}{Theorem}
\newtheorem{nproposition}{Proposition}
\newtheorem{nlemma}{Lemma}

\newtheorem{ndefinition}{Definition}

\theoremstyle{remark}
\newtheorem{nexample}{Example}

\theoremstyle{plain}

\newcommand{\tn}{\textnormal}

\newcommand{\fml}[1]{{\mathcal{#1}}}

\definecolor{gray}{rgb}{.4,.4,.4}
\definecolor{midgrey}{rgb}{0.5,0.5,0.5}
\definecolor{middarkgrey}{rgb}{0.35,0.35,0.35}
\definecolor{darkgrey}{rgb}{0.3,0.3,0.3}
\definecolor{darkred}{rgb}{0.7,0.1,0.1}
\definecolor{midblue}{rgb}{0.2,0.2,0.7}
\definecolor{darkblue}{rgb}{0.1,0.1,0.5}
\definecolor{defseagreen}{cmyk}{0.69,0,0.50,0}

\newcommand{\jnoteF}[1]{}

\newcounter{Comment}[Comment]
\DeclareMathOperator*{\xupgets}{\vdash_{\scriptsize{1}}}

\DeclareMathSymbol{\Delta}{\mathalpha}{operators}{1}
\DeclareMathSymbol{\Theta}{\mathalpha}{operators}{2}
\DeclareMathSymbol{\Pi}{\mathalpha}{operators}{5}
\DeclareMathSymbol{\Sigma}{\mathalpha}{operators}{6}

\newboolean{extended}      
\setboolean{extended}{false} %
\newcommand{\iflongpaper}[1]{\ifthenelse{\boolean{extended}}{#1}{}}
\newcommand{\ifregularpaper}[1]{\ifthenelse{\boolean{extended}}{}{#1}}

\newboolean{addthanks}      
\setboolean{addthanks}{false} %
\newcommand{\ifaddack}[1]{\ifthenelse{\boolean{addthanks}}{#1}{}}

{\noindent \textit{Proof sketch#1.}}{\mbox{}\nobreak\hfill\hspace{6pt}$\Box$}

\newcommand{\mxsat}{MaxSAT\xspace}

\newcommand{\hmxsat}{HornMaxSAT\xspace}

\newcommand{\php}{PHP\xspace}

\newcommand{\urq}{URQ\xspace}

\newcommand{\hencode}{\textsf{HEnc}}

\newcommand{\atmost}{\textsf{AtMost}}
\newcommand{\atleast}{\textsf{AtLeast}}

\newcolumntype{L}[1]{>{\raggedright\let\newline\\\arraybackslash\hspace{0pt}}m{#1}}
\newcolumntype{C}[1]{>{\centering\let\newline\\\arraybackslash\hspace{0pt}}m{#1}}
\newcolumntype{R}[1]{>{\raggedleft\let\newline\\\arraybackslash\hspace{0pt}}m{#1}}

\newtcbox{\dashedbox}[1][]{
  math upper,
  baseline=0.4\baselineskip,
  equal height group=dashedbox,
  nobeforeafter,
  colback=white,
  boxrule=0pt,
  enhanced jigsaw,
  boxsep=0.5pt,
  top=2pt,
  bottom=2pt,
  left=2pt,
  right=2pt,
  borderline horizontal={0.5pt}{0pt}{dashed},
  borderline vertical={0.5pt}{0pt}{dashed},
  #1
}

\newtcbox{\filledbox}[1][]{
  math upper,
  baseline=0.4\baselineskip,
  equal height group=dashedbox,
  nobeforeafter,
  colback=white,
  boxrule=0pt,
  enhanced jigsaw,
  boxsep=0.5pt,
  top=2pt,
  bottom=2pt,
  left=2pt,
  right=2pt,
  borderline horizontal={0.5pt}{0pt}{black},
  borderline vertical={0.5pt}{0pt}{black},
  #1
}

\newcommand{\mailtodomain}[1]{\href{mailto:#1@ciencias.ulisboa.pt}{\nolinkurl{#1}}}

\graphicspath{{plots/w-wbo/}}
\DeclareGraphicsExtensions{.pdf}

\def\jpms{Joao Marques-Silva}
\def\alexi{Alexey Ignatiev}
\def\ajrm{Antonio Morgado}

\title{On Tackling the Limits of Resolution in SAT Solving
\thanks{
  This work was supported by FCT funding of post-doctoral grants
  SFRH/BPD/103609/2014, SFRH/BPD/120315/2016, and LASIGE Research
  Unit, ref. UID/CEC/00408/2013.}
}
\titlerunning{On Tackling the Limits of Resolution in SAT Solving}

\author{
  {\alexi}\inst{1,2}
  \and
  {\ajrm}\inst{1}
  \and
  {\jpms}\inst{1}
}
\authorrunning{Ignatiev, Morgado \& Marques-Silva}

\institute{%
  LASIGE, Faculty of Science, University of Lisbon, Portugal\\
  {{\{\mailtodomain{aignatiev}\texttt{,}\mailtodomain{ajmorgado}\texttt{,}\mailtodomain{jpms}\}\texttt{@ciencias.ulisboa.pt}}}\\
  \and
  ISDCT SB RAS, Irkutsk, Russia
}

\begin {document}
\maketitle
\setcounter{footnote}{0}

%
%
%
%
\begin{abstract}
  The practical success of Boolean Satisfiability (SAT) solvers
  stems from the CDCL (Conflict-Driven Clause Learning) approach to
  SAT solving. However, from a propositional proof complexity
  perspective, CDCL is no more powerful than the resolution proof
  system, for which many hard examples exist.
  This paper proposes a new problem transformation, which enables
  reducing the decision problem for formulas in conjunctive normal
  form (CNF) to the problem of solving maximum satisfiability over
  Horn formulas.
  Given the new transformation, the paper proves a polynomial bound on 
  the number of \mxsat resolution steps for pigeonhole formulas. This
  result is in clear contrast with earlier results on the length of
  proofs of \mxsat resolution for pigeonhole formulas. The paper also
  establishes the same polynomial bound in the case of modern
  core-guided \mxsat solvers.
  Experimental results, obtained on CNF formulas known to be hard for
  CDCL SAT solvers, show that these can be efficiently solved with
  modern MaxSAT solvers.
\end{abstract}
%

%
%
%
%

\section{Introduction} \label{sec:intro}

Boolean Satisfiability (SAT) solvers have made remarkable progress
over the last two decades. Unable to solve formulas with more than a few
hundred variables in the early 90s, SAT solvers are now capable of
routinely solving formulas with a few million
variables~\cite{sat-handbook09,biere-jsat08}. The success of SAT
solvers is supported by the CDCL (Conflict-Driven Clause
Learning)~\cite[Chapter~04]{sat-handbook09} paradigm, and the ability
of SAT solvers to learn clauses from induced
conflicts~\cite{sat-handbook09}.
Nevertheless, being no more powerful than the general resolution
proof system~\cite{darwiche-aij11}, CDCL SAT solvers are also known
not to scale for specific formulas, which are hard for 
resolution~\cite{haken-tcs85,urquhart-jacm87,chvatal-jacm88}.
Recent work has considered different forms of extending CDCL with
techniques adapted from more powerful proof
systems as well as
others~\cite{simon-ijai01,soos-sat09,huang-aij10,simon-aaai10,leberre-jsat10,biere-sat14},
with success in some settings.
Nevertheless, for pigeonhole formulas~\cite{cook-jsl79}, and with the
exception of the lingeling SAT solver~\cite{biere-satcomp13} on
specific encodings, modern CDCL SAT solvers are unable to prove
unsatisfiability even for a fairly small numbers of pigeons.

This paper proposes an alternative path to tackle the difficulties of
the resolution proof system, by developing an approach that aims to
complement existing SAT solvers, and which also builds upon efficient
CDCL SAT solvers. The motivation is to transform the original problem,
from one clausal form to another, the latter enconding a restricted
maximum satisfiability problem, but in such a way that CDCL SAT
solvers can still be exploited.

Given any CNF formula $\fml{F}$, the paper shows how to encode the
problem as Horn Maximum Satisfiability (\hmxsat), more concretely by
requiring a given cost on the \hmxsat formulation. This enables
solving the modified problem with either a MaxSAT solver or with a
dedicated \hmxsat solver.
The basic encoding is also shown to be subject to a number of
optimizations, that can effectively reduce the number of variables.
%
The paper then shows that for propositional encodings of the
pigeonhole principle~\cite{cook-jsl79}, transformed to \hmxsat,
there exists a polynomially time bounded sequence of \mxsat resolution
steps which enables deriving a number of falsified clauses that 
suffices for proving unsatisfiable the original \php formula.
Similarly, the paper also proves that for modern core-guided \mxsat
solvers there exist sequences of unsatisfiable cores that enable
reaching the same conclusion in polynomial time.
This in turn suggests that MaxSAT algorithms~\cite{mhlpms-cj13} can be
effective in practice when applied to such instances.

Experimental results, obtained on different encodings of the
pigeonhole principle, but also on other instances that are well-known to be
hard for resolution~\cite{urquhart-jacm87}, confirm the theoretical
result. Furthermore, a recently-proposed family of MaxSAT
solvers~\cite{bacchus-cp11,jarvisalo-sat16}, based on iterative
computation of minimum hitting sets, is also shown to be effective in
practice and on a wider range of classes of instances.

The paper is organized as follows. \autoref{sec:prelim} introduces the
definitions and notation used throughout the
paper. \autoref{sec:sat2hmxsat} develops a simple encoding from SAT
into \hmxsat. \autoref{sec:proofs} derives a polynomial bound on the
number and size of \mxsat-resolution steps to establish the
unsatisfiability of propositional formulas encoding the pigeonhole
principle transformed into \hmxsat. The section also shows that there
are executions of core-guided \mxsat solvers that take polynomial time
to establish a lower bound of the cost of the \mxsat solution which
establishes the unsatisfiability of the original CNF formula.
Experimental results on formulas encoding the pigeonhole principle,
but also on other formulas known to be hard for CDCL SAT
solvers~\cite{urquhart-jacm87} are analyzed in~\autoref{sec:res}.
The paper concludes in~\autoref{sec:conc}.

\jnoteF{Use no more than 2 pages in total: abstract+introduction.}


%
%
%

\section{Preliminaries} \label{sec:prelim}

The paper assumes definitions and notation standard in propositional
satisfiability (SAT) and maximum satisfiability
(\mxsat)~\cite{sat-handbook09}.
Propositional variables are taken from a set $X=\{x_1,x_2,\ldots\}$.
A Conjunctive Normal Form (CNF) formula is defined as a conjunction of
disjunctions of literals, where a literal is a variable or its
complement. CNF formulas can also be viewed as sets of sets of
literals, and are represented with calligraphic letters, $\fml{A}$,
$\fml{F}$, $\fml{H}$, etc.
A truth assignment is a map from variables to $\{0,1\}$. Given a
truth assignment, a clause is satisfied if at least one of its literals
is assigned value 1; otherwise it is falsified. A formula is satisfied
if all of its clauses are satisfied; otherwise it is falsified.
If there exists no assignment that satisfies a CNF formula $\fml{F}$,
then $\fml{F}$ is referred to as \emph{unsatisfiable}.
(Boolean) Satisfiability (SAT) is the decision problem for
propositional formulas, i.e.\ to decide whether a given propositional
formula is satisfiable.
Since the paper only considers propositional formulas in CNF,
throughout the paper SAT refers to the decision problem for
propositional formulas in CNF.

To simplify modeling with propositional logic, one often represents
more expressive constraints. Concrete examples are cardinality
constraints and pseudo-Boolean constraints~\cite{sat-handbook09}.
A cardinality constraint of the form $\sum x_i\le k$ is referred to as
an $\atmost{}k$ constraint, whereas a cardinality constraint of the
form $\sum x_i\ge k$ is referred to as an $\atleast{}k$ constraint.
The study of propositional encodings of cardinality and pseudo-Boolean
constraints is an area of active
research~\cite{warners-ipl98,bailleux-cp03,sinz-cp05,een-jsat06,sat-handbook09,nieuwenhuis-sat09,codish-lpar10,roussel-sat09,nieuwenhuis-cj11,nieuwenhuis-sat11,koshimura-ictai13a}.

%
A clause is Horn if it contains at most one positive literal. A Horn
clause is a \emph{goal} clause if it has no positive literals;
otherwise it is a \emph{definite} clause.
The decision problem for Horn formulas is well-known to be in P, with
linear time algorithms since the
80s~\cite{gallier-jlp84,minoux-ipl88}.
A number of function problems defined on Horn formulas can be solved
in polynomial time~\cite{msimp-jelia16}. These include computing the
lean kernel, finding a minimal unsatisfiable subformula and finding
a maximal satisfiable subformula.

\subsection{Propositional Encodings of the Pigeonhole Principle}

The propositional encoding of the pigeonhole hole principle is
well-known~\cite{cook-jsl79}.
\begin{ndefinition}[Pigeonhole Principle, \php~\cite{cook-jsl79}]
  The pigeonhole principle states that if $m+1$ pigeons are
  distributed by $m$ holes, then at least one hole contains more than
  one pigeon. A more formal formulation is that  there exists no
  injective function mapping from $\{1,2,...,m+1\}$ to
  $\{1,2,...,m\}$, for $m\ge 1$.
\end{ndefinition}
Propositional formulations of $\text{\php}$ encode the negation of the
principle, and ask for an assignment such that the $m+1$ pigeons are
placed into $m$ holes.
The propositional encoding of the $\text{\php}^{m+1}_{m}$ problem can
be derived as follows. Let the variables be $x_{ij}$, with $1\le i\le
m+1, 1\le j\le m$, with $x_{ij}=1$ iff the $i^{\text{th}}$ pigeon is
placed in the $j^{\text{th}}$ hole.
The constraints are that each pigeon must be placed in at least one
hole, and each hole must not have more than one pigeon.
\begin{equation}
  \bigwedge_{i=1}^{m+1} \atleast1(x_{i1},\ldots,x_{im}) \land
  \bigwedge_{j=1}^{m} \atmost1(x_{1j},\ldots,x_{m+1,j})\\
\end{equation}
An $\atleast1$ constraint can be encoded with a single clause. For the
$\atmost1$ constraint there are different encodings,
including~\cite{sat-handbook09,sinz-cp05,een-jsat06}.
For example, the pairwise encoding~\cite{sat-handbook09} of
$\atmost1(x_{1j},\ldots,x_{m+1,j})$ uses no auxiliary variables and
the clauses $\land_{r=2}^{m+1}\land_{s=1}^{r-1}(\neg x_{rj}\lor\neg x_{sj})$.
It is well-known that resolution has an exponential lower bound for
$\text{\php}$~\cite{haken-tcs85,beame-focs96,razborov-dlt01}.

\subsection{\mxsat, \mxsat Resolution \& \mxsat Algorithms} \label{ssec:mxsat}

\paragraph{\mxsat.}
For unsatisfiable formulas, the maximum satisfiability (\mxsat)
problem is to find an assignment that maximizes the number of
satisfied clauses (given that not all clauses can be satisfied).
There are different variants of the \mxsat
problem~\cite[Chapter~19]{sat-handbook09}. Partial \mxsat allows for
\emph{hard} clauses (which must be satisfied) and \emph{soft} clauses
(which represent a preference to satisfy those clauses). There are
also weighted variants, in which soft clauses are given a weight, and
for which hard clauses (if any) have a weight of $\top$ (meaning
clauses that must be satisfied).
The notation $(c, w)$ will be used to represent a clause $c$ with $w$
denoting the cost of falsifying $c$. The paper considers partial
MaxSAT instances, with hard clauses, for which $w=\top$, and soft
clauses, for which $w=1$. The notation $\langle\fml{H},\fml{S}\rangle$
is used to denote partial \mxsat problems with sets of hard
($\fml{H}$) and soft ($\fml{S}$) clauses.
Throughout the paper, a \mxsat \emph{solution} represents either a
maximum cardinality set of satisfied soft clauses or an assignment that
satisfies all hard clauses and also maximizes (or minimizes, resp.) the
number of satisfied (or falsified, resp.) soft clauses.

\paragraph{\mxsat Resolution~\cite{bonet-aij07,larrosa-aij08}.}
In contrast with SAT, the \mxsat resolution operation requires the
introduction of additional clauses other than the resolvent, and
resolved clauses cannot be resolved again.
Let $(x\lor A, u)$ and $(\neg x\lor B, w)$ be two clauses, and let
$m\triangleq\min(u,w)$,
$u\ominus w\triangleq(u\;\text{==}\,\top)\,\text{?}\,\top:u-w$, with
$u\ge w$.
The (non-clausal) \mxsat resolution step~\cite{larrosa-aij08} is shown
in~\autoref{tab:mxres}. (We could have used the clausal
formulation~\cite{bonet-aij07}, but it is more verbose and unnecessary
for the purposes of the paper. It suffices to mention that clausal
\mxsat resolution adds at most $2n$ clauses at each resolution step,
where the number of variables is $n$ and the number of literals in
each clause does not exceed $n$.)
\begin{table}[t]
  \begin{center}
    \caption{Example \mxsat-resolution steps.}\label{tab:mxres}
    \renewcommand{\tabcolsep}{0.35em}
    \scalebox{0.975}{
      \begin{tabular}{C{1.75cm}C{1.75cm}C{8cm}} \toprule
        Clause 1 & Clause 2 & Derived Clauses \\ \midrule
        $(x\lor A, u)$ &
        $(\neg x\lor B, w)$ &
        $(A\lor B, m)$,
        $(x\lor A, u\ominus m)$,
        $(\neg x\lor B, w\ominus m)$, \newline
        $(x\lor A\lor\neg B, m)$,
        $(\neg x\lor\neg A\lor B, m)$
        \\ \midrule
        $(x\lor A, 1)$ &
        $(\neg x, \top)$ &
        $(A, 1)$,
        $(\neg x, \top)$, 
        $(\neg x\lor\neg A, 1)$
        \\ \bottomrule
      \end{tabular}
    }
  \end{center}
\end{table}
It is well-known that \mxsat-resolution is unlikely to improve
propositional resolution~\cite{bonet-aij07}. For the original
$\text{\php}^{m+1}_{m}$ formulas, there are known exponential lower bounds on
the size of deriving one empty clause by \mxsat-resolution (given that
the remaining clauses are
satisfiable)~\cite[Corollary~18]{bonet-aij07}.

\paragraph{\mxsat Algorithms.}
Many algorithms for \mxsat have been proposed over the
years~\cite[Chapter~19]{sat-handbook09}.
The most widely investigated can be broadly organized into branch and
bound~\cite[Chapter~19]{sat-handbook09},
iterative-search~\cite{malik-sat06,leberre-jsat10,koshimura-jsat12},
core-guided~\cite{malik-sat06,msp-corr07,mhlpms-cj13,ansotegui-aij13,narodytska-aaai14,mdms-cp14,lynce-cp14},
and minimum hitting sets~\cite{bacchus-cp11,jarvisalo-sat16}.
In most proposed algorithms, core-guided and minimum hitting sets
\mxsat algorithms iteratively determine formulas to be unsatisfiable,
until satisfiability is reached for a formula that relaxes clauses of
minimum cost.
This paper analyzes the operation of core-guided \mxsat algorithms,
concretely the MSU3 algorithm~\cite{msp-corr07}~\footnote{Different
  implementations of the MSU3 have been proposed over the
  years~\cite{msp-corr07,mhlpms-cj13,ansotegui-aij13,lynce-cp14},
  which often integrate different improvements. A well-known
  implementation of MSU3 is OpenWBO~\cite{lynce-cp14}, one of the best
  \mxsat solvers in the \mxsat Evaluations since 2014.}. 
Moreover, and to our best knowledge, the relationship between
core-guided \mxsat algorithms and \mxsat resolution was first
investigated in~\cite{narodytska-aaai14}.

\subsection{Related Work} \label{ssec:relw}

The complexity of resolution on pigeonhole formulas has been studied
by different authors,
e.g. see~\cite{cook-jsl79,haken-tcs85,beame-focs96,razborov-dlt01,nordstrom-siglog15}
and references therein, among others.
It is well-known that for other proof systems, including cutting
planes and extended resolution, PHP has polynomial
proofs~\cite{cook-sigact76,buss-jsl87,turan-dam87,buss-tcs88,atserias-cp04,segerlind-bsl07}.
Different authors have looked into extending CDCL (and so
resolution) with the goal of solving formulas for which resolution has
known exponential lower bounds~\cite{goldberg-cade02,goldberg-amai05,soos-sat09,huang-aij10,simon-aaai10,leberre-jsat10,biere-jsat08,moura-cade11,moura-jar13,biere-sat14}.
Some SAT solvers apply pattern matching
techniques~\cite{biere-satcomp13}, but these are only effective for
specific propositional encodings. Furthermore, there has been limited
success in applying cutting planes and extended resolution in
practical SAT solvers.

\jnoteF{Use no more than 2 pages in total:
  preliminaries+MaxSAT algorithms+MaxSAT resolution.}


%
%
%


%
%
%

\section{Reducing SAT to HornMaxSAT} \label{sec:sat2hmxsat}

The propositional satisfiability problem for CNF formulas can be
reduced to \hmxsat, more concretely to the problem of deciding whether
for some target Horn formula there exists an assignment that satisfies
a given number of soft clauses.

Let $\fml{F}$ be a CNF formula, with $N$ variables $\{x_1\ldots,x_N\}$
and $M$ clauses $\{c_1,\ldots,c_M\}$.
Given $\fml{F}$, the reduction creates a Horn \mxsat problem with hard
clauses $\fml{H}$ and soft clauses $\fml{S}$,
$\langle\fml{H},\fml{S}\rangle=\hencode(\fml{F})$.
For each variable $x_i\in X$, create new variables $p_i$ and $n_i$,
where $p_i=1$ iff $x_i=1$, and $n_i=1$ iff $x_i=0$. Thus, we need a hard
clause $(\neg p_i\lor\neg n_i)$, to ensure that we do not
simultaneously assign $x_i=1$ and $x_i=0$. (Observe that the added
clause is Horn.)
This set of hard Horn clauses is referred to as $\fml{P}$.
For each clause $c_j$, we require $c_j$ to be satisfied, by requiring
that one of its literals \emph{not} to be falsified. For each literal
$x_i$ use $\neg n_i$, and for each literal $\neg x_i$ use $\neg p_i$.
Thus, $c_j$ is encoded with a new (hard) clause $c'_j$  with the same
number of literals as $c_j$, but with only negative literals on the
$p_i$  and $n_i$ variables, and so the resulting clause is also Horn. 
The set of soft clauses $\fml{S}$ is given by $(p_i)$ and $(n_i)$ for
each of the original variables $x_i$.
If the resulting Horn formula has a \hmxsat solution with at least
$N$ variables assigned value 1, then the original formula is
satisfiable; otherwise the original formula is unsatisfiable. (Observe
that, by construction, the \hmxsat solution cannot assign value 1 to
more than $N$ variables. Thus, unsatisfiability implies being unable
to satisfy more than $N-1$ soft clauses.)
Clearly, the encoding outlined in this section can be the subject of
different improvements, e.g.\ not all clauses need to be goal clauses.
(An approach to tighten the encoding is detailed later in this section.)

\begin{nexample}
  Let the CNF formula be:
  \begin{equation}
    (x_1\lor\neg x_2\lor x_3)\land(x_2\lor x_3)\land(\neg x_1\lor\neg x_3)
  \end{equation}
  The new variables are $\{n_1,p_1,n_2,p_2,n_3,p_3\}$. Preventing
  simultaneous assignment to 0 and 1 is guaranteed with the hard
  clauses:
  \begin{equation}
    (\neg n_1\lor\neg p_1)\land(\neg n_2\lor\neg p_2)\land(\neg n_3\lor\neg p_3)
  \end{equation}
  The original clauses are reencoded as hard clauses as follows:
  \begin{equation}
    (\neg n_1\lor\neg p_2\lor\neg n_3)\land(\neg n_2\lor\neg n_3)\land(\neg p_1\lor\neg p_3)
  \end{equation}
  Finally, the soft clauses are
  $\fml{S}=\{(n_1),(p_1),(n_2),(p_2),(n_3),(p_3)\}$.
\end{nexample}

The transformation proposed above can be related with the well-known 
dual-rail encoding, used in different
settings~\cite{BryantBBCS87,mfmso-ictai97,RoordaC05,jmsss-jelia14,pimms-ijcai15}. To
our best knowledge, the use of a dual-rail encoding for deriving a
pure Horn formula has not been proposed in earlier work.

\begin{nlemma} \label{lm:nub}
  Given $\langle\fml{H},\fml{S}\rangle=\hencode(\fml{F})$, there can
  be no more than $N$ satisfied soft clauses.
\end{nlemma}
\begin{proof}
  By construction of $\langle\fml{H},\fml{S}\rangle$, for any $x_i$, 
  there is no assignment that satisfies $\fml{H}$ with $n_i=1$ and
  $p_i=1$. 
\end{proof}

\begin{nlemma} \label{lm:f2h}
  Let $\fml{F}$ have a satisfying assignment $\nu$. Then, there exists
  an assignment that satisfies $\fml{H}$ and $N$ soft clauses in
  $\langle\fml{H},\fml{S}\rangle=\hencode(\fml{F})$.
\end{nlemma}

\begin{proof}
  Given $\nu$, we create an assignment $\nu'$ to the $n_i$ and $p_i$
  variables that satisfies the clauses in $\fml{H}$, and $N$ clauses
  in $\fml{S}$.
  For each $x_i$, if $\nu(x_i)=1$, then set $p_i=1$; otherwise set
  $n_i=1$. Thus, there will be $N$ satisfied clauses in $\fml{S}$.
  For each clause $c_j\in\fml{F}$, pick a literal $l_k$ assigned value
  1. If $l_k=x_k$, then $c'_k$ contains literal $\neg n_k$, and so it
  is satisfied. If $l_k=\neg x_k$, then $c'_k$ contains literal $\neg
  p_k$, and so it is satisfied.
  Thus every clause in $\fml{H}$ is satisfied, and $N$ soft clauses
  are satisfied.
\end{proof}

\begin{nlemma} \label{lm:h2f}
  Let $\nu'$ be an assignment that satisfies the clauses in $\fml{H}$
  and $N$ clauses in $\fml{S}$. Then there exists an assignment $\nu$
  that satisfies $\fml{F}$. 
\end{nlemma}

\begin{proof}
  By construction of $\langle\fml{H},\fml{S}\rangle$, for each $x_i$,
  either $n_i$ is assigned value 1, or $p_i$ is assigned value 1, but
  not both. Let $\nu(x_i)=1$ if $\nu'(p_i)=1$ and $\nu(x_i)=0$ if
  $\nu'(n_i)=1$. All variables $x_i$ are either assigned value 0 or 1.
  For clause $c'_j$, let $l_k$ be a literal assigned value 1. If
  $l_k=\neg n_k$, then $x_k$ is a literal in $c_j$ and since
  $\nu(x_i)=1$, then the clause $c_j$ is satisfied. Otherwise, if
  $l_k=\neg p_k$, then $\neg x_k$ is a literal in $c_j$ and since
  $\nu(x_i)=0$, then the clause $c_j$ is satisfied.
\end{proof}

\autoref{lm:nub},~\autoref{lm:f2h} and~\autoref{lm:h2f} yield the
following.

\begin{ntheorem} \label{thm:cnf2horn}
  $\fml{F}$ is satisfiable if and only if there exists an assignment
  that satisfies $\fml{H}$ and $N$ clauses in $\fml{S}$.
\end{ntheorem}

The reduction of SAT into \hmxsat can also be applied to the
$\text{PHP}^{m+1}_m$ problem.

\begin{nexample}[Pigeonhole Principle]
  With each variable $x_{ij}$, $1\le i\le m+1, 1\le j\le m$, we
  associate two new variables: $n_{ij}$ and $p_{ij}$. The set of
  clauses $\fml{P}$ prevents a variable $x_i$ from being assigned
  value 0 and 1 simultaneously:
  $\fml{P}=\{(\neg n_{ij}\lor\neg p_{ij})\,|\,1\le i\le m+1, 1\le j\le m\}$.
  $\fml{L}_i$ represents the encoding of each $\atleast1$ constraint,
  concretely $\fml{L}_i=(\neg n_{i1}\lor\ldots\lor\neg n_{im})$.
  $\fml{M}_j$ represents the encoding of each $\atmost1$ constraint,
  which will depend on the encoding used.
  The soft clauses $\fml{S}$ are given by,
  \[\small
  \begin{split}
    \{ & (n_{11}),\ldots,(n_{1m}),\ldots,(n_{m+1\,1}),\ldots,(n_{m+1\,m}),\\
    & (p_{11}),\ldots,(p_{1m}),\ldots,(p_{m+1\,1}),\ldots,(p_{m+1\,m}) \}
  \end{split}
  \]
  with $|\fml{S}|=2m(m+1)$. 
  Thus, the complete reduction of PHP into \mxsat becomes:
  \begin{equation} \label{eq:mphp}
    \hencode\left(\text{\php}^{m+1}_{m}\right)
    \triangleq\langle\fml{H},\fml{S}\rangle=\left\langle
    \land_{i=1}^{m+1}\fml{L}_i\land
    \land_{j=1}^{m}\fml{M}_j\land\fml{P},
    \fml{S}\right\rangle
  \end{equation}
  Clearly, given $\fml{P}$, one cannot satisfy more the $m(m+1)$
  soft clauses. By~\autoref{thm:cnf2horn}, $\text{\php}^{m+1}_m$ is
  satisfiable if and only if there exists an assignment that satisfies
  the hard clauses $\fml{H}$ and $m(m+1)$ soft clauses from
  $\fml{S}$. 
\end{nexample}

\paragraph{Reducing the number of variables.}
As hinted earlier in this section, one can devise encodings that use
fewer variables, while producing a Horn \mxsat formula.
Given $\fml{F}$, let $\fml{N}\subseteq\fml{F}$ denote the set of
non-Horn clauses.
For each clause $c_j\in\fml{N}$, let $C_j$ represent the set of
positive literals of $c_j$, and let $x_i\in C_j$.
Moreover, let $W=\cup_j C_j$, i.e.\ the set of variables causing the
clauses in $\fml{N}$ not to be Horn.
Finally, let $r_i=1$ if and only if for clause $c_j$ variable $x_i$ is
not dual-rail encoded. Clearly, for $c_j$ we must have $\sum_{x_i\in
  C_j}r_i\le1$, i.e.\ the number of positive literals that remain
cannot exceed 1; otherwise the resulting clause would not be Horn.
Moreover, we would like to discard as many variables as possible, and
so we add a soft clause $(r_i)$ for each $x_i\in W$. This corresponds to
finding a maximum independent set, which we can approximate
heuristically. 

\jnoteF{We can add an example if necessary, or remove the paragraph in
  case of lack of space.}

\jnoteF{Use no more than 2.5 pages in total:
  reduction+comment on dual-rail encoding+simple example+PHP example+correctness.}


%
%
%

\section{Short \mxsat Proofs for PHP} \label{sec:proofs}

This section shows that the reduction of $\text{\php}^{m+1}_{m}$ to \hmxsat
based on a dual-rail encoding enables both existing core-guided \mxsat
algorithms and also \mxsat resolution, to prove in polynomial time that the
original problem formulation\footnote{This section studies the original
  \emph{pairwise} encoding of $\text{\php}^{m+1}_m$. However, a similar
  argument can be applied to $\text{\php}^{m+1}_m$ provided any encoding of
  AtMost1 constraints $\fml{M}_j$, as confirmed by the experimental results in
\autoref{sec:res}.} is unsatisfiable.
Recall from~\autoref{thm:cnf2horn}, that $\text{\php}^{m+1}_{m}$ is satisfiable
if and only if, given~\eqref{eq:mphp}, there exists an assignment that
satisfies $\fml{H}$ and $m(m+1)$ soft clauses in $\fml{S}$.
This section shows that for both core-guided algorithms and for \mxsat
resolution, we can conclude in polynomial time that satisfying $\fml{H}$
requires falsifying at least $m(m+1)+1$ soft clauses, thus proving
$\text{\php}^{m+1}_{m}$ to be unsatisfiable.

The results in this section should be contrasted with earlier
work~\cite{bonet-aij07}, which proves that \mxsat resolution requires
an exponentially large proof to produce an empty clause, this assuming
the \emph{original} propositional encoding for $\text{\php}^{m+1}_{m}$.

\subsection{A Polynomial Bound on Core-Guided \mxsat
  Algorithms} \label{ssec:cg-proof}

This section shows that a core-guided \mxsat algorithm will conclude
in polynomial time that more than $m(m+1)$ clauses must be falsified,
when the hard clauses are satisfied, thus proving the original
$\text{\php}^{m+1}_{m}$ to be unsatisfiable. The analysis assumes the
operation of basic core-guided algorithm, MSU3~\cite{msp-corr07}, but
similar analyses could be carried out for other families of
core-guided algorithms\footnote{Basic knowledge of core-guided \mxsat
  algorithms is assumed. The reader is referred to recent surveys for
  more information~\cite{mhlpms-cj13,ansotegui-aij13}.}.

The following observations about~\eqref{eq:mphp} are essential to prove
the bound on the run time.
First, the clauses in the $\fml{L}_i$ constraints do not share
variables in common with the clauses in the $\fml{M}_j$ constraints.
Second, each constraint $\fml{L}_i$ is of the form $(\neg
n_{i1}\lor\ldots\lor\neg n_{im})$ and so its variables are disjoint from
any other $\fml{L}_k$, $k\not=i$.
Third, assuming a pairwise encoding, each constraint $\fml{M}_j$ is of
the form $\land_{r=2}^{m+1}\land_{s=1}^{r-1}(\neg p_{rj}\lor\neg
p_{sj})$, and so its variables are disjoint from any other
$\fml{M}_l$, $l\not=j$.
Since the sets of variables for each constraint are disjoint from the
other sets of variables, we can exploit this partition of the clauses,
and run a \mxsat solver \emph{separately} on each one. (Alternatively,
we could assume the MSU3 \mxsat algorithm to work with disjoint
unsatisfiable cores.)

\autoref{tab:cg-proof} summarizes the sequence of unit propagation
steps that yields a lower bound on the number of falsified clauses
larger than $m(m+1)$.
\begin{table}[t]
  \caption{Partitioned core-guided unit propagation steps.} \label{tab:cg-proof}
  \begin{adjustbox}{center}
    \renewcommand{\tabcolsep}{0.35em}
    \scalebox{0.825}{
      \begin{tabular}{C{1.5cm}C{3.25cm}C{2.35cm}C{2.35cm}C{2.575cm}C{1.3cm}}
        \toprule
        Constraint & Hard clause(s) & Soft clause(s) & Relaxed clauses
        & Updated AtMost$k$ Constraints & LB \newline increase\\
        \midrule
        $\fml{L}_i$ &
        $(\neg n_{i1}\lor\ldots\lor\neg n_{im})$ &
        $(n_{i1}),\ldots,(n_{im})$\newline &
        $(r_{il}\lor n_{i1}),$\newline$1\le l\le m$ &
        $\sum_{l=1}^{m} r_{il}\le 1$ &
        1
        \\ \midrule
        $\fml{M}_j$ &
        $(\neg p_{1j}\lor\neg p_{2j})$ &
        $(p_{1j}),(p_{2j})$ &
        $(r_{1j}\lor p_{1j})$, $(r_{2j}\lor p_{2j})$ &
        $\sum_{l=1}^{2}r_{lj}\le 1$ &
        1
        \\ \midrule
        $\fml{M}_j$ &
        $(\neg p_{1j}\lor\neg p_{3j})$,\newline
        $(\neg p_{2j}\lor\neg p_{3j})$,\newline
        $(r_{1j}\lor p_{1j})$, \newline
        $(r_{2j}\lor p_{2j})$, \newline
        $\sum_{l=1}^{2}r_{lj}\le 1$ &
        $(p_{3j})$ &
        $(r_{3j}\lor p_{3j})$ &
        $\sum_{l=1}^{3}r_{lj}\le 2$ &
        1
        \\ \midrule
        \multicolumn{6}{c}{$\cdots$}
        \\ \midrule
        $\fml{M}_j$ &
        $(\neg p_{1j}\lor\neg p_{m+1j}), \ldots$,\newline
        $(\neg p_{mj}\lor\neg p_{m+1j})$,\newline
        $(r_{1j}\lor p_{1j}), \ldots$, \newline
        $(r_{mj}\lor p_{mj})$, \newline
        $\sum_{l=1}^{m}r_{lj}\le m-1$ &
        $(p_{m+1j})$ &
        $(r_{m+1j}\lor p_{m+1j})$ &
        $\sum_{l=1}^{m+1}r_{lj}\le m$ &
        1
        \\ \bottomrule
      \end{tabular}
    }
  \end{adjustbox}
\end{table}
For each $\fml{L}_i$, the operation is summarized in the second row
of~\autoref{tab:cg-proof}. Unit propagation yields a conflict between
$m$ soft clauses and the corresponding hard clause. This means that
at least one of these soft clauses must be falsified.
Since there are $m+1$ constraints $\fml{L}_i$, defined on disjoint
sets of variables, then each will contribute at least one falsified
soft clause, which puts the lower bound on the number of falsified
clauses at $m+1$.

For each $\fml{M}_j$ the operation is summarized in rows 3 to last
of~\autoref{tab:cg-proof}. Each row indicates a sequence of unit
propagation steps that produces a conflict, each on a distinct set of
soft clauses.
Observe that each soft clause $(p_{kj})$, $k\ge2$, induces a sequence of
unit propagation steps, that causes the AtMost$\{k-1\}$ constraint to
become inconsistent.
Concretely, for iteration $k$ (where row 3 corresponds to
iteration~1), the sequence of unit propagation steps is summarized
in~\autoref{tab:cg-proof2}~\footnote{The notation $\Phi\xupgets\bot$
  indicates that inconsistency (i.e.\ a falsified clause) is derived
  by unit propagation on the propositional encoding of $\Phi$.
  This is the case with existing encodings of \atmost$k$ constraints.
  %
}.
\begin{table}[t]
  \begin{center}
    \caption{Analysis of $\fml{M}_j$, iteration $k$.} \label{tab:cg-proof2}
    \renewcommand{\tabcolsep}{0.5em}
    \scalebox{0.975}{
      \begin{tabular}{L{6.0cm}L{4.0cm}} \toprule
        Clauses & Unit Propagation \\ \midrule
        $(p_{k+1\,j})$ & $p_{k+1\,j} = 1$ \\ \midrule
        $(\neg p_{1j}\lor\neg p_{k+1\,j}),\ldots,(\neg p_{kj}\lor\neg
        p_{k+1\,j})$ & $p_{1j}=\ldots=p_{kj}=0$ \\ \midrule
        $(r_{1j}\lor p_{1j}),\ldots,(r_{kj}\lor p_{kj})$ &
        $r_{1j}=\ldots=r_{kj}=1$ \\ \midrule
        $\sum_{l=1}^{k}r_{lj}\le k-1$ &
        $\left(\sum_{l=1}^{k}r_{lj}\le k-1\right)\xupgets\bot$\\ \bottomrule
      \end{tabular}
    }
  \end{center}
\end{table}
Since there are $m$ such rows, then each $\fml{M}_j$ contributes at
least $m$ falsified soft clauses. Moreover, the number of $\fml{M}_j$
constraints is $m$, and so the $\fml{M}_j$ constraints increase the
bound by $m\cdot m$.

Given the above, in total we are guaranteed to falsify at least
$m+1+m\cdot m = m(m+1)+1$ clauses, thus proving that one cannot
satisfy $m(m+1)$ soft clauses if the hard clauses are satisfied.
In turn, this proves that the $\text{\php}^{m+1}_{m}$ problem is
unsatisfiable.

We can also measure the run time of the sequence of unit propagation
steps. For each $\fml{L}_i$, the run time is $\fml{O}(m)$, and
there will be $m$ such unit propagation steps, for a total
$\fml{O}(m^2)$.
For each $\fml{M}_j$ there will be $m$ unit propagation steps, with
run time between $\fml{O}(1)$ and $\fml{O}(m)$. Thus, the run time of
the sequence of unit propagation steps for each $\fml{M}_j$ is
$\fml{O}(m^2)$. Since there are $m$ constraints $\fml{M}_j$, then the
total run time is~$\fml{O}(m^3)$.

\begin{nproposition}
  Given~\eqref{eq:mphp}, and for a core-guided MSU3-like \mxsat
  solver, there is a sequence of unit propagation steps such that a
  lower bound of $m(m+1)+1$ is computed in $\fml{O}(m^3)$ time.
\end{nproposition}

\begin{proof}(Sketch)
  The discussion above.
\end{proof}

Moreover, it is important to observe that the unit propagation steps
considered in the analysis above avoid the clauses in $\fml{P}$,
i.e.\ only the clauses in $\fml{L}_i$, $\fml{M}_j$, $\fml{S}$, and
relaxed clauses, are used for deriving the lower bound of $m(m+1)+1$
on the minimum number of falsified soft clauses. As shown
in~\autoref{sec:res}, and for the concrete case of \php, the clauses
in $\fml{P}$ are unnecessary and actually impact negatively the
performance of core-guided \mxsat solvers.
Finally, and although the proof above assumes an MSU3-like core-guided
algorithm, similar ideas could be considered in the case of other
variants of core-guided \mxsat
algorithms~\cite{malik-sat06,mhlpms-cj13,ansotegui-aij13,narodytska-aaai14}.

\subsection{A Polynomial Bound on \mxsat
  Resolution} \label{ssec:mxsatres-proof}

We can now exploit the intuition from the previous section to identify
the sequence of \mxsat resolution steps that enable deriving
$m(m+1)+1$ empty clauses, thereby proving that \emph{any} assignment
that satisfies the hard clauses must falsify at least $m(m+1)+1$ soft
clauses, and therefore proving that the propositional encoding of \php
is unsatisfiable. As before, we assume that the pairwise encoding is
used to encode each constraint $\fml{M}_j$.
As indicated earlier in~\autoref{ssec:mxsat}, we consider a simplified
version of \mxsat resolution~\cite{larrosa-aij08}, which is
non-clausal. As explained below, this is not problematic, as just a
few clauses are of interest. For the clausal version of \mxsat
resolution, the other clauses, which our analysis ignores, are
guaranteed to be linear in the number of variables at each step, and
will \emph{not} be considered again.

\autoref{tab:mr-proof} summarizes the essential aspects of the \mxsat
resolution steps used to derive $m(m+1)+1$ empty
clauses. (Also,~\autoref{ssec:cg-proof} clarifies that the formula can
be partitioned if $\fml{P}$ is ignored.)
Similarly to the previous section, the $\fml{L}_i$ constraints serve
to derive $m+1$ empty clauses, whereas each $\fml{M}_j$ constraint
serves to derive $m$ empty clauses. In total, we derive $m(m+1)+1$
empty clauses, getting the intended result.
\begin{table}[t]
  \caption{Simplified \mxsat resolution steps.} \label{tab:mr-proof}
  \begin{adjustbox}{center}
    \scalebox{0.925}{
      \renewcommand{\tabcolsep}{0.35em}
      \begin{tabular}{C{1.5cm}C{3.75cm}C{7cm}} \toprule
        Constraint & Clauses & Resulting clause(s)\\
        \midrule
        $\fml{L}_i$ &
        $(\neg n_{i1}\lor\ldots\lor\neg n_{im},\top),$\newline $(n_{i1},1)$
        &
        $\dashedbox{(\neg n_{i2}\lor\ldots\lor\neg n_{im}, 1)}$, $\ldots$
        \\ \midrule
        $\fml{L}_i$ &
        $(\neg n_{i2}\lor\ldots\lor\neg n_{im},1),$\newline $(n_{i2},1)$
        &
        $\dashedbox{(\neg n_{i3}\lor\ldots\lor\neg n_{im}, 1)}$, $\ldots$
        \\ \midrule
        \multicolumn{3}{c}{$\cdots$}
        \\ \midrule
        $\fml{L}_i$ &
        $(\neg n_{im},1),$\newline $(n_{im},1)$
        &
        $\filledbox[colback=lightgray]{(\bot, 1)}\,$, $\ldots$
        \\ \midrule
        $\fml{M}_j$ &
        $(\neg p_{1j}\lor\neg p_{2j},\top),$\newline $(p_{1j},1)$
        &
        $(\neg p_{2j}, 1)$,
        $(\neg p_{1j}\lor\neg p_{2j}, \top)$,
        $\dashedbox{(p_{1j}\lor p_{2j}, 1)}$
        \\ \midrule
        $\fml{M}_j$ &
        $(\neg p_{2j},1),$\newline $(p_{2j},1)$
        &
        $\filledbox[colback=lightgray]{(\bot, 1)}\,$
        \\ \midrule
        $\fml{M}_j$ &
        $(\neg p_{1j}\lor\neg p_{3j},\top),$\newline $(p_{1j}\lor p_{2j},1)$
        &
        $\dashedbox{(p_{2j}\lor\neg p_{3j},1)}$,
        $(\neg p_{1j}\lor\neg p_{3j},\top)$,
        $(\neg p_{1j}\lor\neg p_{3j}\lor\neg p_{2j},1)$,
        $\dashedbox{(p_{1j}\lor p_{2j}\lor p_{3j}, 1)}\,$
        \\ \midrule
        $\fml{M}_j$ &
        $(\neg p_{2j}\lor\neg p_{3j},\top),$\newline $(p_{2j}\lor\neg p_{3j},1)$
        &
        $\dashedbox{(\neg p_{3j},1)}$,
        $(\neg p_{2j}\lor\neg p_{3j},\top)$
        \\ \midrule
        $\fml{M}_j$ &
        $(\neg p_{3j},1),$\newline $(p_{3j},1)$
        &
        $\filledbox[colback=lightgray]{(\bot, 1)}\,$
        \\ \midrule
        \multicolumn{3}{c}{$\cdots$}
        \\ \midrule
        $\fml{M}_j$ &
        $(\neg p_{1j}\lor\neg p_{m+1j},\top),$\newline
        $(p_{1j}\lor\ldots\lor p_{mj}, 1)$
        &
        $\dashedbox{(p_{2j}\ldots p_{mj}\lor\neg p_{m+1j}, 1)}\,$, $\ldots$
        \\ \midrule
        $\fml{M}_j$ &
        $(\neg p_{2j}\lor\neg p_{m+1j},\top),$\newline
        $(p_{2j}\lor\ldots\lor p_{mj}\lor\neg p_{m+1j}, 1)$
        &
        $\dashedbox{(p_{3j}\ldots p_{mj}\lor\neg p_{m+1j}, 1)}\,$, $\ldots$
        \\ \midrule
        \multicolumn{3}{c}{$\cdots$}
        \\ \midrule
        $\fml{M}_j$ &
        $(\neg p_{mj}\lor\neg p_{m+1j},\top),$\newline
        $(p_{mj}\lor\neg p_{m+1j}, 1)$
        &
        $\dashedbox{\neg p_{m+1j}, 1)}\,$, $\ldots$
        \\ \midrule
        $\fml{M}_j$ &
        $(p_{m+1j},1),$\newline
        $(\neg p_{m+1j}, 1)$
        &
        $\filledbox[colback=lightgray]{(\bot, 1)}\,$
        \\ \bottomrule
      \end{tabular}
    }
    \end{adjustbox}
\end{table}
As shown in \autoref{tab:mr-proof}, for each constraint $\fml{L}_i$,
start by applying \mxsat resolution between the hard clause
$\fml{L}_i\triangleq(\neg n_{i1}\lor\ldots\lor\neg n_{im})$ and soft
clause $(n_{i1})$ to get soft clause $(\neg n_{i2}\lor\ldots\lor\neg
n_{im})$, and a few other clauses (which are irrelevant for our
purposes).
Next, apply $m-1$ additional \mxsat resolution steps, resolving soft
clause $(\neg n_{ik}\lor\ldots\lor\neg  n_{im})$ with soft clause $(n_{ik})$
to get soft clause $(\neg n_{ik+1}\lor\ldots\lor\neg  n_{im})$.
Clearly, the final \mxsat resolution step will yield an empty
clause. Therefore, over all $m+1$ $\fml{L}_i$ constraints, we derive
$m+1$ empty clauses.

\autoref{tab:mr-proof} also illustrates the application of the
\mxsat resolution steps to the pairwise encoding of $\fml{M}_j$.
At iteration $i$, with $2\le i\le m+1$, we apply $i$ \mxsat resolution
steps to derive another empty clause. In total, we derive $m$ empty
clauses for each $\fml{M}_j$.
An essential aspect is selecting the initial clause from which each
sequence of \mxsat resolution steps is executed. These reused clauses
are highlighted in~\autoref{tab:mr-proof}, and are crucial for getting
the right sequence of \mxsat resolution steps.
For each $\fml{M}_j$, the \mxsat resolution steps can be organized in
$m$ phases, each yielding an empty soft clause. For phase $l$, the
previous phase $l-1$ produces the clause
$(p_{1j}\lor p_{2j}\lor\ldots\lor p_{lj},1)$, which is then
iteratively simplified, using unit soft clauses, until the empty soft
clause for phase $l$ is derived. It should be noted that the first
phase uses two unit soft clauses to produce $(p_{1j}\lor p_{2j},1)$,
which is then used in the second phase. As in~\autoref{ssec:cg-proof},
is is immediate that each soft clause is \emph{never} reused.

Regarding the run time complexity, observe that each \mxsat resolution
step runs in time linear on the number of literals in the clauses. The
clauses in the problem formulation have no more than $\fml{O}(m)$
literals. This also holds true as \mxsat resolution steps are applied.
By analogy with the analysis of the core-guided algorithm, a total of
$\fml{O}(m^2)$ empty soft clauses will be derived. From the analysis
above, summarized in~\autoref{tab:mr-proof}, deriving the
$\fml{O}(m^2)$ empty clauses requires a total of $\fml{O}(m^3)$ \mxsat
resolution steps. For non-clausal \mxsat resolution, since the number
of generated (non-clausal) terms is constant for each \mxsat
resolution step, then the run time is $\fml{O}(m^3)$. In contrast, for
clausal \mxsat resolution~\cite[Definition~1]{bonet-aij07}, since the
number of literals for each resolution step is $\fml{O}(m^2)$, then
the run time becomes $\fml{O}(m^5)$.

\begin{nproposition}
  For the \hmxsat encoding of $\tn{\php}^{m+1}_{m}$, there exists a
  polynomial sequence of \mxsat resolution steps, each producing a
  number of constraints polynomial in the size of the problem
  formulation, that produces $m(m+1)+1$ soft empty clauses.
\end{nproposition}

\begin{proof}(Sketch)
  The discussion above.
\end{proof}

\subsection{Integration in SAT Solvers}

\jnoteF{Use no more than 0.5+1+2+0.5=4 pages in total:
  motivation+core-guided+mxsat resolution+integration.}
\jnoteF{Alexey: we need to integrate your comments!}

This section shows that off-the-shelf \mxsat solvers, which build on
CDCL SAT solvers, can solve $\text{\php}^{m+1}_{m}$ in polynomial
time, provided the right order of conflicts is chosen. In turn, this
motivates integrating core-guided \mxsat reasoning into SAT solvers.
Similarly, one could consider integrating \mxsat resolution (or a mix
of both~\cite{narodytska-aaai14}) but, like resolution, \mxsat
resolution is harder to implement in practice.
The proposed problem transformation can be applied on demand, and the
operation of CDCL can be modified to integrate some form of
core-guided reasoning.
In contrast to other attempts at extending CDCL, the use of \mxsat
reasoning, will build also on CDCL itself.

\paragraph{MaxHS-like Horn \mxsat.}
The reduction to Horn \mxsat also motivates the development of
dedicated \mxsat solvers. One approach is to build upon
MaxHS-solvers~\cite{bacchus-cp11,jarvisalo-sat16}, since in this case
the SAT checks can be made to run in linear time, e.g.\ using an
implementation of LTUR~\cite{minoux-ipl88}.
As indicated above, similar technique can possibly be integrated into
SAT solvers.

\paragraph{Handling $\fml{P}$ clauses.}
The $\fml{P}$ clauses prevent assigning a variable simultaneously
value 0 and value 1. As the analysis for the PHP instances suggests,
and the experimental results confirm, these clauses can be responsible
for non-polynomial run times.
One can envision attempting to solve problems without considering the
$\fml{P}$ clauses, and then adding these clauses on demand, as deemed
necessary to block non-solutions. The operation is similar to the
well-known counterexample-guided abstraction refinement paradigm
(CEGAR)~\cite{clarke-jacm03}.


%
%
%

\section{Experimental Evaluation} \label{sec:eval}

This section evaluates the ideas proposed in the paper in practice,
for the case of formulas that are known to be hard for
resolution-based reasoning.
Concretely, the experimental evaluation shows that the performance
gains are provided by the proposed problem transformation and the
follow-up core-guided MaxSAT solving.

\subsection{Experimental Setup} \label{sec:setup}

To illustrate the main points of the paper, a number of solvers were tested.
However and in order to save space, the results are detailed below only for
some of the tested competitors.\footnote{The discussion focuses on the results
  of the best performing \emph{representatives} of the considered families of
  solvers. Solvers that are missing in the discussion are meant to be
``dominated'' by their representatives, i.e.\ these solve fewer instances.}
The families of the evaluated solvers as well as the chosen representatives for
the families are listed in \autoref{tab:solvers}.
The family of CDCL SAT solvers comprises MiniSat~2.2 (\emph{minisat})
and Glucose~3 (\emph{glucose}) while the family of SAT solvers
strengthened with the use of other powerful techniques (e.g.\ Gaussian
elimination, GA and/or cardinality-based reasoning, CBR) includes
lingeling (\emph{lgl}) and CryptoMiniSat (\emph{crypto}).
The MaxSAT solvers include the known tools based on implicit minimum-size
hitting set enumeration, i.e.\ MaxHS (\emph{maxhs}) and LMHS (\emph{lmhs}), and
also a number of core-guided solvers shown to be best for industrial instances
in a series of recent MaxSAT
Evaluations\footnote{\url{http://www.maxsat.udl.cat}}, e.g.\ MSCG
(\emph{mscg}), OpenWBO16 (\emph{wbo}) and WPM3 (\emph{wpm3}), as well as the
recent MaxSAT solver Eva500a (\emph{eva}) based on MaxSAT resolution.
Other competitors considered include CPLEX (\emph{lp}), OPB solvers
cdcl-cuttingplanes (\emph{cc}) and Sat4j (\emph{sat4j}) as well as a solver
based on ZBDDs called ZRes (\emph{zres}).

\begin{table*}[!t]
  \caption{Families of solvers considered in the evaluation (their best
  performing representatives are written in \emph{italics}). \emph{SAT+}
  stands for SAT strengthened with other techniques, \emph{IHS MaxSAT} is
  for implicit hitting set based MaxSAT, \emph{CG MaxSAT} is for core-guided
  MaxSAT, \emph{MRes} is for MaxSAT resolution, \emph{MIP} is for mixed integer
  programming, \emph{OPB} is for pseudo-Boolean optimization, \emph{BDD}
  is for binary decision diagrams.}
  \scriptsize
  \begin{center}
    \begin{tabular}{ccccccccccccccc}
      \toprule
      \multicolumn{2}{c}{{\footnotesize \textbf{SAT}}} &
      \multicolumn{2}{c}{{\footnotesize \textbf{SAT+}}} &
      \multicolumn{2}{c}{{\footnotesize \textbf{IHS MaxSAT}}} &
      \multicolumn{3}{c}{{\footnotesize \textbf{CG MaxSAT}}} &
      {\footnotesize \textbf{MRes}} &
      {\footnotesize \textbf{MIP}} &
      \multicolumn{2}{c}{{\footnotesize \textbf{OPB}}} &
      {\footnotesize \textbf{BDD}} \\
      \cmidrule(lr){1-2}
      \cmidrule(lr){3-4}
      \cmidrule(lr){5-6}
      \cmidrule(lr){7-9}
      \cmidrule(lr){10-10}
      \cmidrule(lr){11-11}
      \cmidrule(lr){12-13}
      \cmidrule(lr){14-14}
      minisat &
      \emph{glucose} &
      \emph{lgl} &
      crypto &
      \emph{maxhs} &
      \emph{lmhs} &
      \emph{mscg} &
      \emph{wbo} &
      wpm3 &
      \emph{eva} &
      \emph{lp} &
      \emph{cc} &
      sat4j &
      \emph{zres} \\
      \midrule
      \cite{een-sat03} &
      \cite{audemard-sat13} &
      \cite{biere-satcomp13,biere-pos14} &
      \cite{soos-sat09,soos-pos10} &
      \cite{bacchus-cp11,davies-sat13,davies-cp13} &
      \cite{jarvisalo-sat16} &
      \cite{mims-jsat15} &
      \cite{martins-sat14} &
      \cite{ansotegui-ijcai15} &
      \cite{narodytska-aaai14} &
      \cite{cplex-page} &
      \cite{elffers-page} &
      \cite{leberre-jsat10} &
      \cite{simon-ijai01} \\
      \bottomrule
    \end{tabular}
  \end{center} \label{tab:solvers}
\end{table*}

Note that three configurations of CPLEX were tested: (1) the default
configuration and the configurations used in (2) MaxHS and (3) LMHS.  Given the
overall performance, we decided to present the results for one best performing
configuration, which turned out to be the default one.
Also, the performance of CPLEX was measured for the following two types of LP
instances: (1) the instances encoded to LP directly from the original CNF
formulas (see \emph{lp-cnf}) and (2) the instances obtained from the \hmxsat
formulas (\emph{lp-wcnf}).
A similar remark can be made with regard to the cc solver: it can deal
with the original CNF formulas as well as their OPB encodings (the
corresponding configurations of the solver are \emph{cc-cnf} and
\emph{cc-opb}\footnote{The two tested versions of \emph{cc-opb}
  (implementing linear and binary search) behave almost identically with a
  minor advantage of linear search. As a result, \emph{cc-opb} stands for
  the linear search version of the solver.}, respectively).

Regarding the IHS-based MaxSAT solvers, both MaxHS and LMHS implement the
\emph{Eq-Seeding} constraints~\cite{davies-sat13}.
Given that all soft clauses constructed by the proposed \hmxsat transformation
are \emph{unit} and that the set of all variables of \hmxsat formulas is
\emph{covered} by the soft clauses, these eq-seeding constraints replicate the
complete MaxSAT formula on the MIP side.
As a result, after all disjoint unsatisfiable cores are enumerated by MaxHS or
LMHS, only one call to an MIP solver is needed to compute the optimum solution.
In order to show the performance of an IHS-based MaxSAT solver with this
feature disabled, we additionally considered another configuration of LMHS
called \emph{lmhs-nes}.\footnote{We chose LMHS (not MaxHS) because it has a
command-line option to disable eq-seeding.}

All the conducted experiments were performed in Ubuntu Linux on an Intel
Xeon~E5-2630 2.60GHz processor with 64GByte of memory. The time limit was set
to 1800s and the memory limit to 10GByte for each individual process to run.

\subsection{Experimental Results} \label{sec:res}

The efficiency of the selected competitors was assessed on the benchmark suite
consisting of 3 sets: (1) pigeonhole formulas (PHP)~\cite{cook-jsl79}, (2)
Urquhart formulas (\emph{URQ})~\cite{urquhart-jacm87}, and (3) their
combinations (\emph{COMB}).

\subsubsection{Pigeonhole Principle benchmarks.}
\begin{figure*}[!t]
  \begin{subfigure}[b]{0.49\textwidth}
    \centering
    \includegraphics[width=\textwidth]{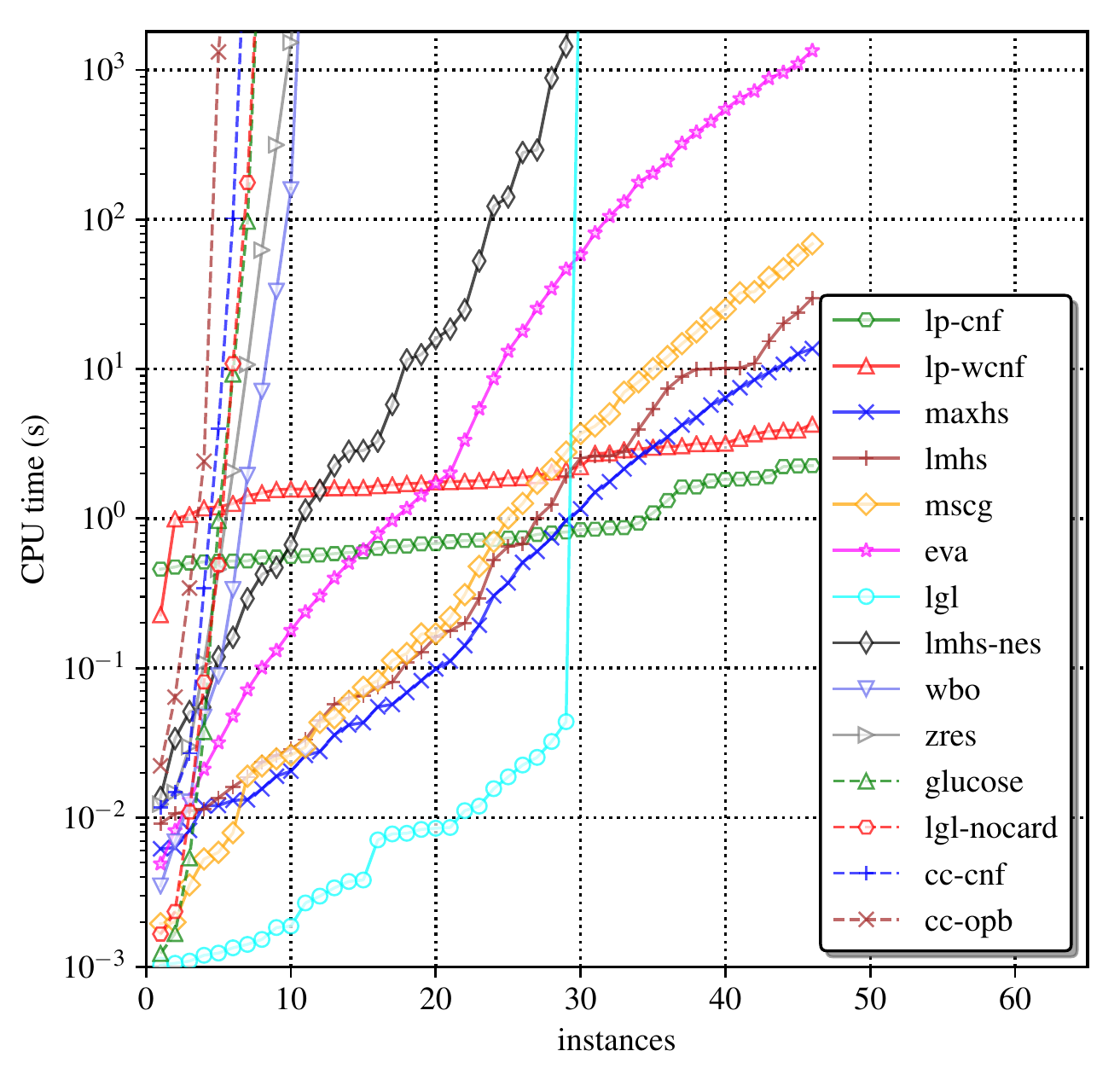}
    \caption{PHP-pw (pairwise)}
    \label{fig:cactus-phppw}
  \end{subfigure}%
  \begin{subfigure}[b]{0.49\textwidth}
    \centering
    \includegraphics[width=\textwidth]{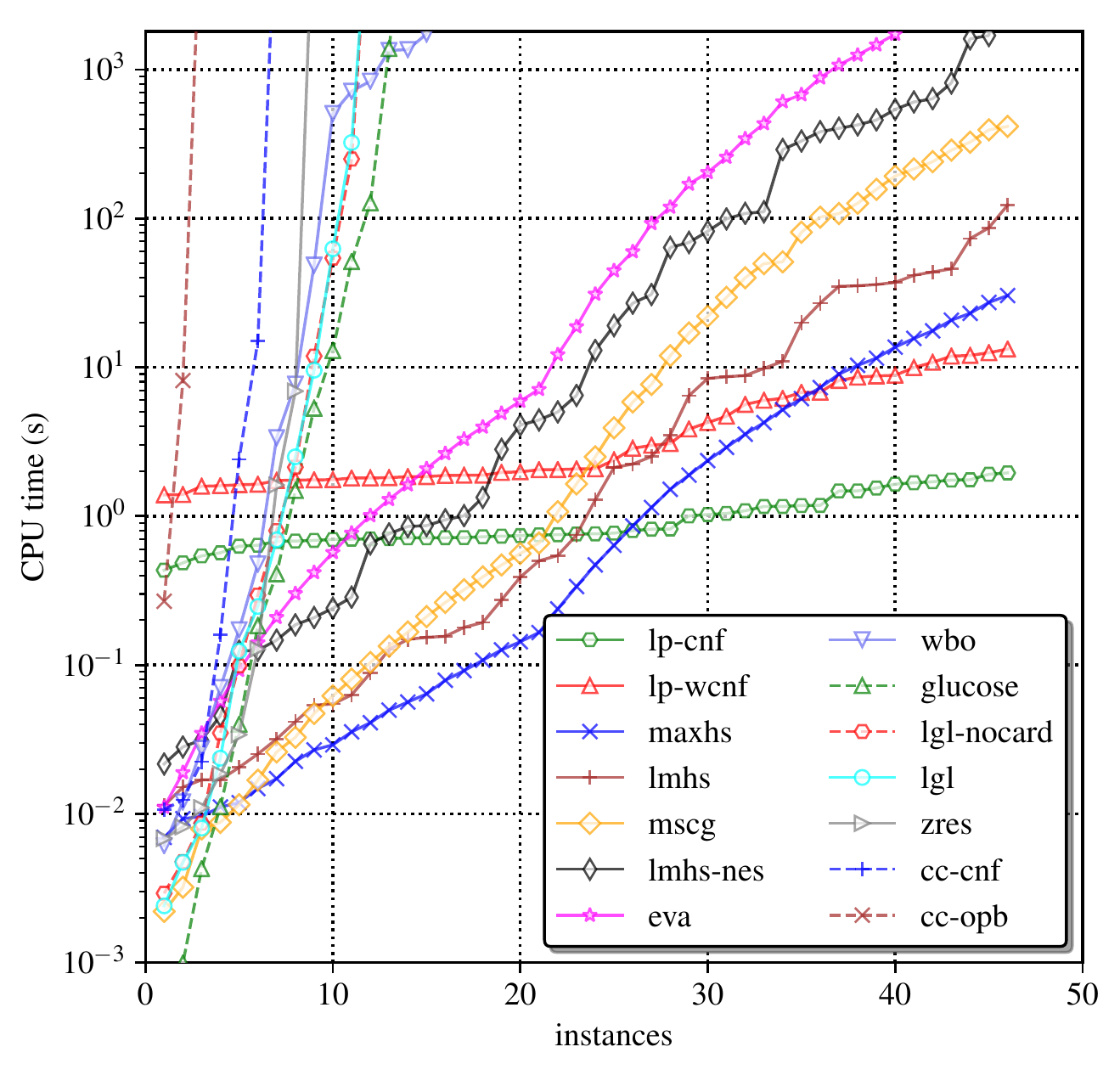}
    \caption{PHP-sc (sequential counter)}
    \label{fig:cactus-phpsc}
  \end{subfigure}
  \caption{Performance of the considered solvers on pigeonhole formulas.}
  \label{fig:php}
\end{figure*}

The set of PHP formulas contains 2 families of benchmarks differing in the way
$\atmost1$ constraints are encoded: (1) standard pairwise-encoded
(\emph{PHP-pw}) and (2) encoded with sequential counters~\cite{sinz-cp05}
(\emph{PHP-sc}).
Each of the families contains 46 CNF formulas encoding the pigeonhole principle
for 5 to 100 pigeons.
\autoref{fig:php}\footnote{Note that all the shown cactus plots below scale the
Y axis logarithmically.} shows the performance of the solver on sets PHP-pw and
PHP-sc. As can be seen, the MaxSAT solvers (except \emph{eva} and \emph{wbo})
and also \emph{lp-$\ast$} are able to solve all instances.  As
expected, CDCL SAT solvers perform poorly for PHP with the exception of
lingeling, which in some cases detects cardinality constraints in
PHP-pw. However, disabling cardinality constraints reasoning or
considering the PHP-sc benchmarks impairs its performance tremendously.
Also note that we were unable to reproduce the performance of \emph{zres}
applied to PHP reported in~\cite{simon-ijai01}.


\subsubsection{On discarding $\fml{P}$ clauses.}
\begin{figure*}[!t]
  \begin{subfigure}[b]{\textwidth}
    \centering
    \includegraphics[width=\textwidth]{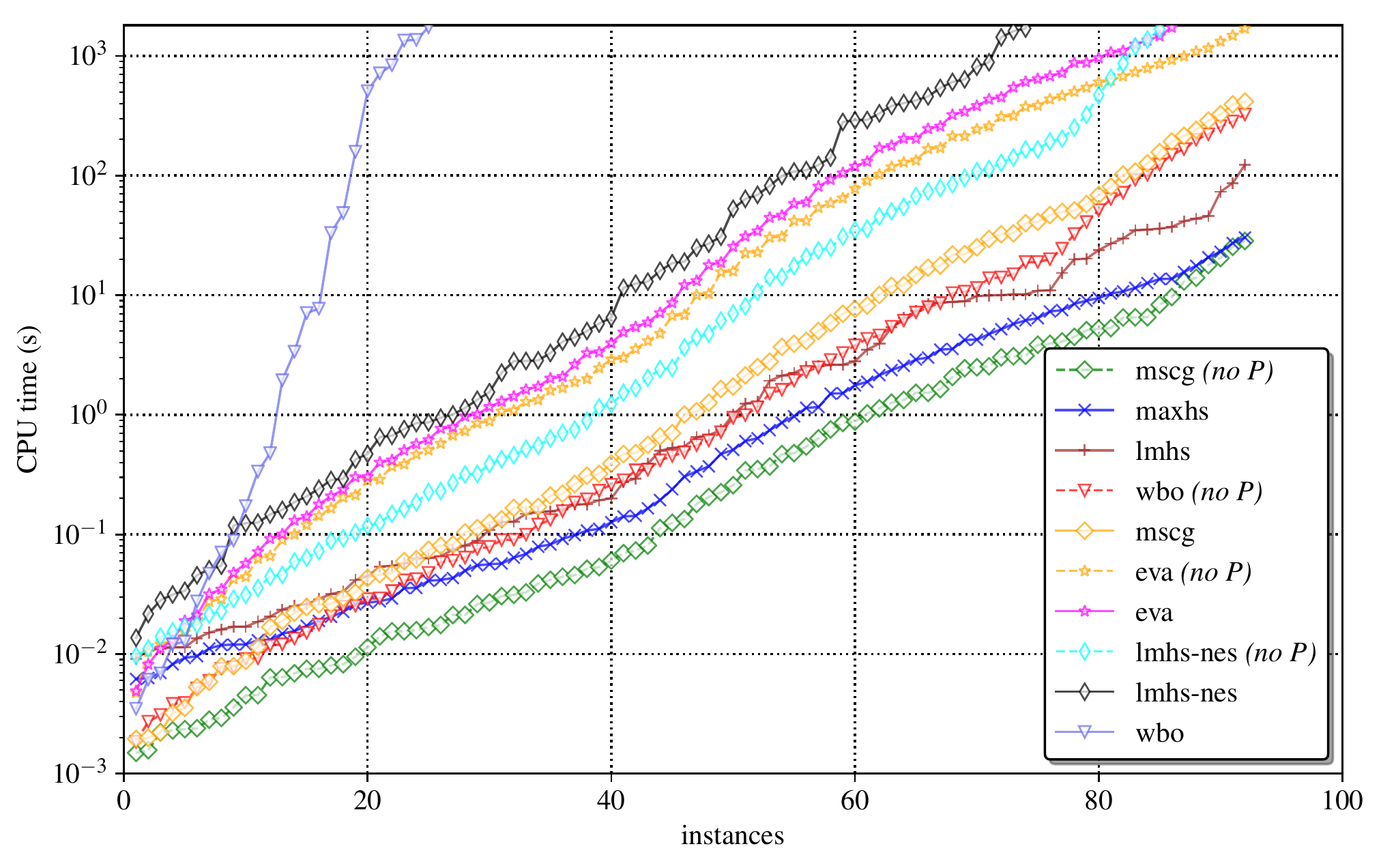}
    \caption{Cactus plot}
    \label{fig:cactus-nop}
  \end{subfigure}

  \begin{subfigure}[b]{0.49\textwidth}
    \centering
    \includegraphics[width=\textwidth]{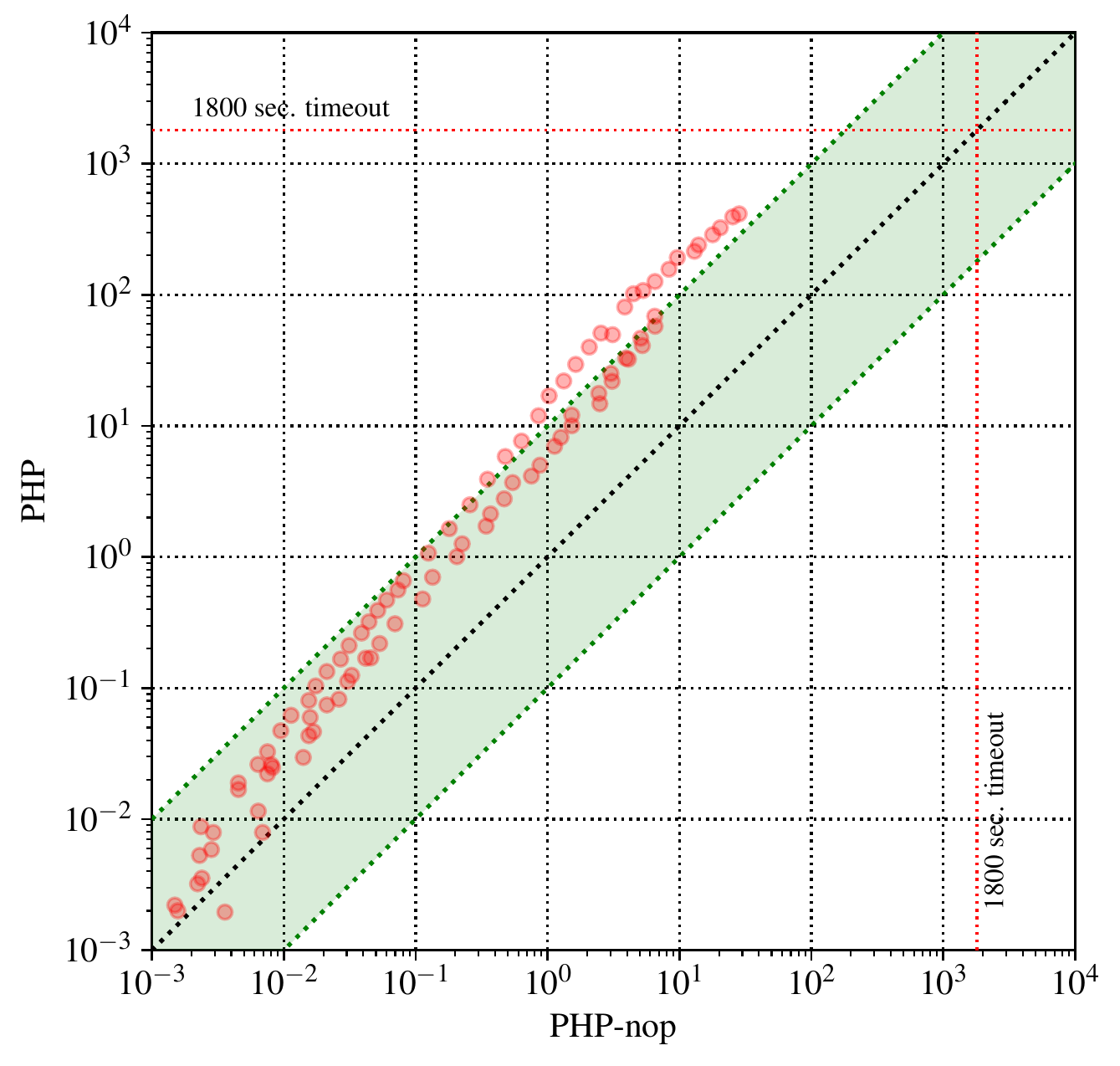}
    \caption{Performance of \emph{mscg} w/ and w/o $\fml{P}$ clauses}
    \label{fig:scatter-nop2}
  \end{subfigure}%
  \begin{subfigure}[b]{0.49\textwidth}
    \centering
    \includegraphics[width=\textwidth]{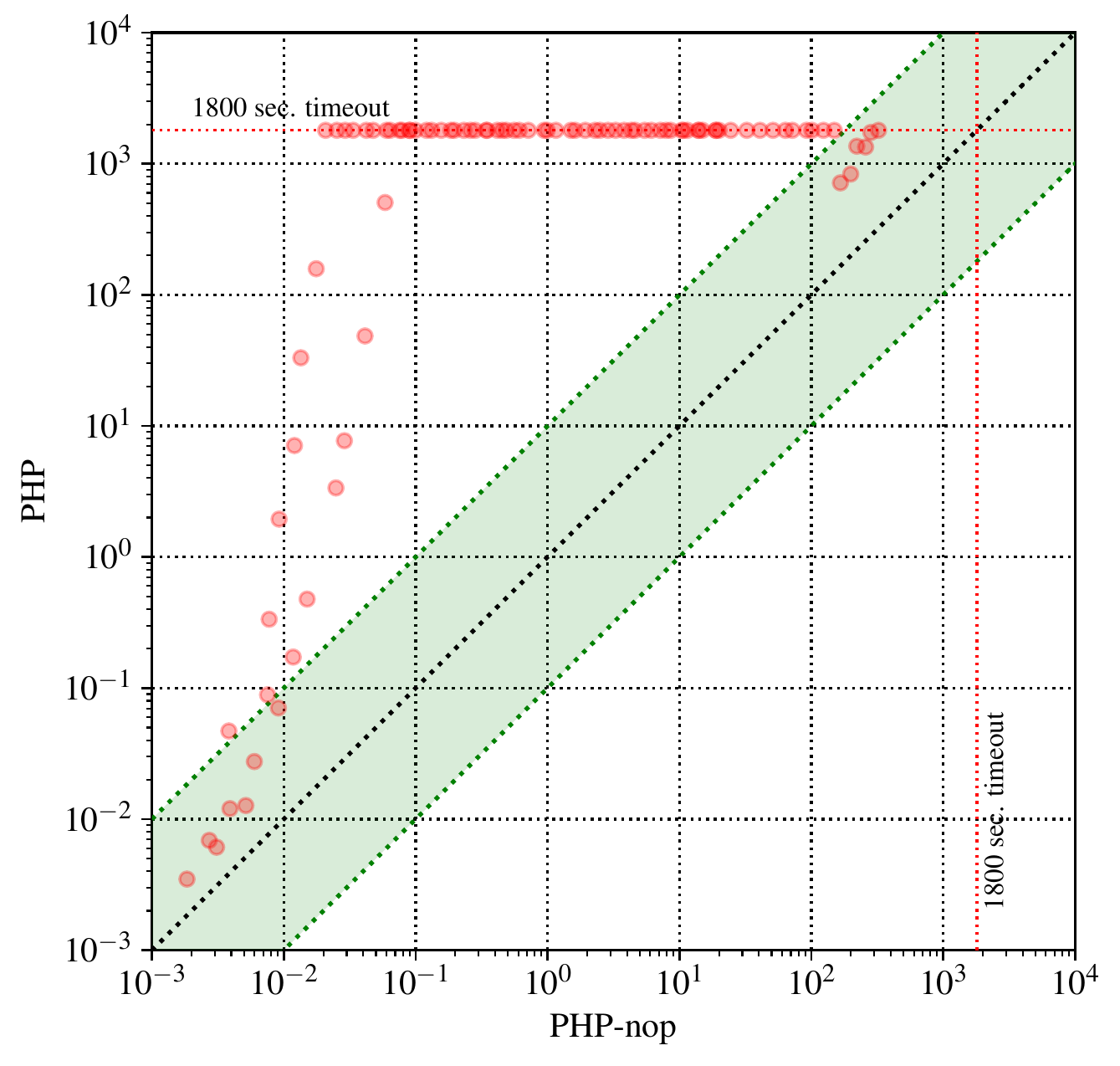}
    \caption{Performance of \emph{wbo} w/ and w/o $\fml{P}$ clauses}
    \label{fig:scatter-nop}
  \end{subfigure}
  \caption{Performance of MaxSAT solvers on PHP-pw $\cup$ PHP-sc w/ and
  w/o $\fml{P}$ clauses.}
  \label{fig:php-nop}
\end{figure*}

To confirm the conjecture that the $\fml{P}$ clauses can hamper a MaxSAT
solver's ability to get \emph{good} unsatisfiable cores, we also considered
both PHP-pw and PHP-sc instances \emph{without} the $\fml{P}$ clauses.
\autoref{fig:php-nop} compares the performance of the MaxSAT solvers working on
PHP formulas w/ and w/o the $\fml{P}$ clauses. The lines with \emph{(no P)}
denote solvers working on the formulas w/o $\fml{P}$ (except \emph{maxhs} and
\emph{lmhs} whose performance is not affected by removal of $\fml{P}$).
As detailed in \autoref{fig:scatter-nop}, the efficiency of \emph{wbo} is
improved by a few orders of magnitude if the $\fml{P}$ clauses are discarded.
Also, as shown in \autoref{fig:scatter-nop2}, \emph{mscg} gets about an order
of magnitude performance improvement outperforming all the other solvers.

\subsubsection{Urquhart benchmarks and combined instances.}

\begin{figure*}[!t]
  \begin{subfigure}[b]{0.49\textwidth}
    \centering
    \includegraphics[width=\textwidth]{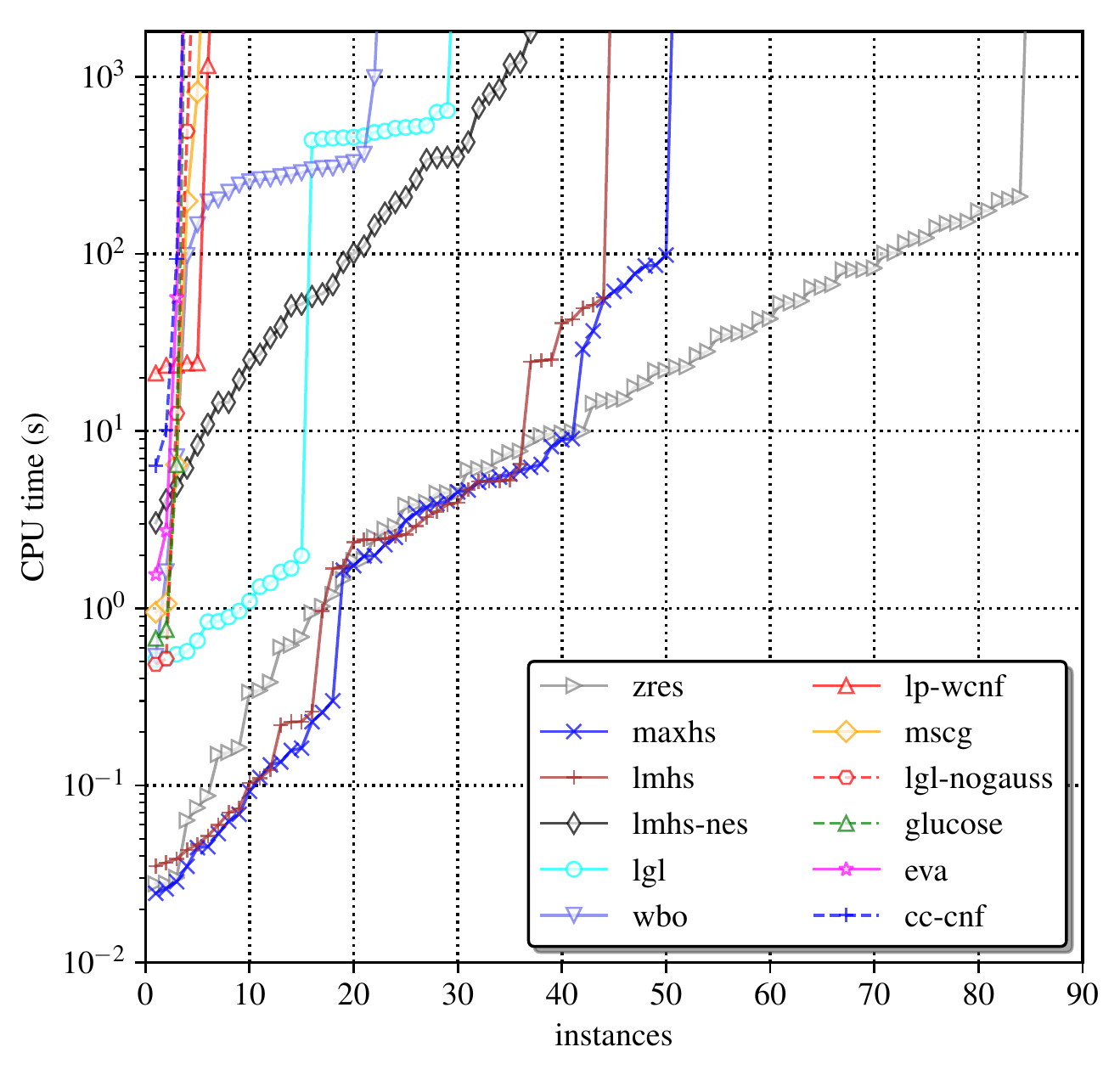}
    \caption{URQ instances}
    \label{fig:cactus-urq}
  \end{subfigure}%
  \begin{subfigure}[b]{0.49\textwidth}
    \centering
    \includegraphics[width=\textwidth]{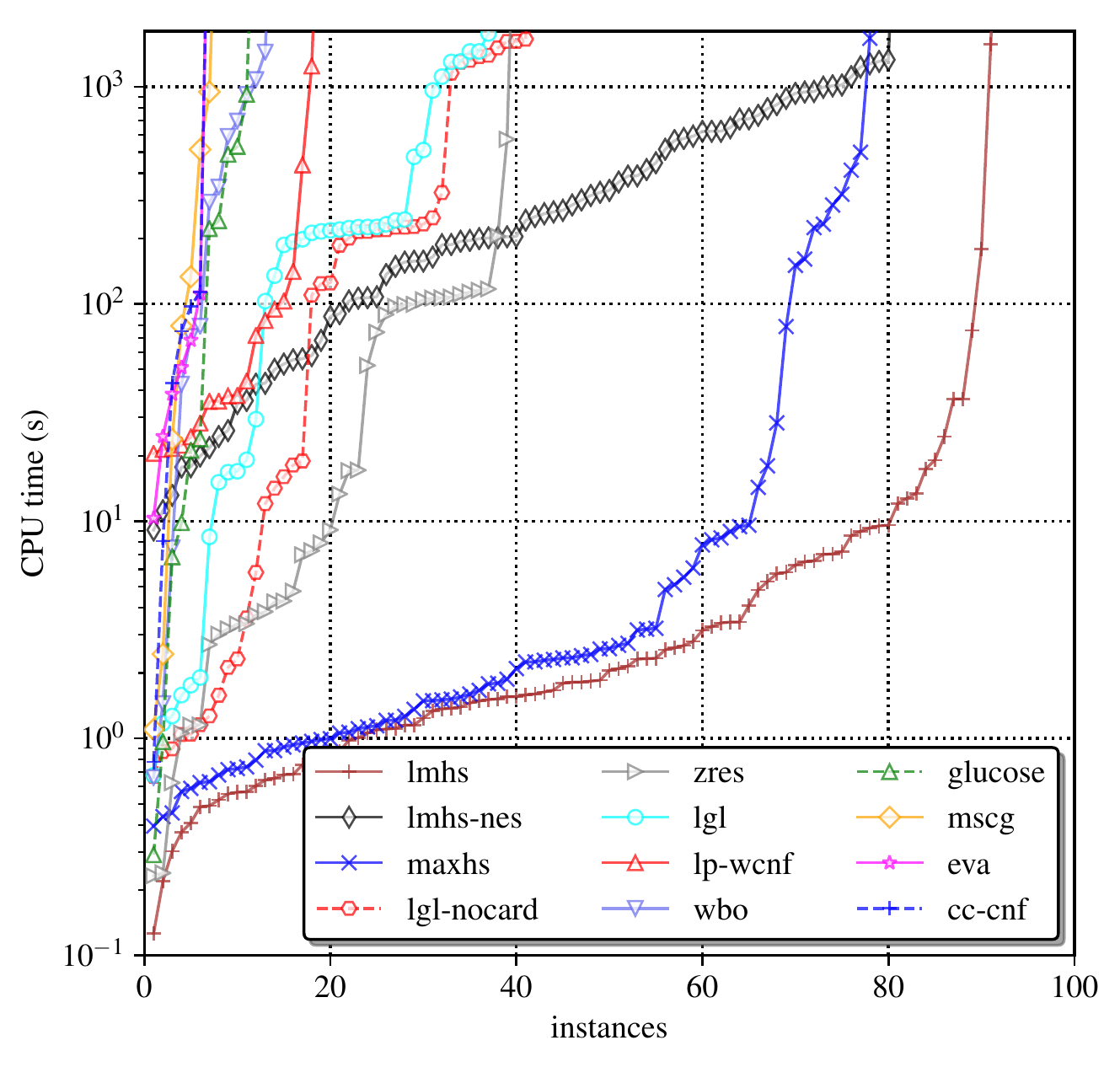}
    \caption{COMB instances}
    \label{fig:cactus-comb}
  \end{subfigure}
  \caption{Performance of the considered solvers on URQ and combined formulas.}
  \label{fig:other}
\end{figure*}

The URQ instances are known to be hard for resolution~\cite{urquhart-jacm87},
but not for BDD-based reasoning~\cite{simon-ijai01}.
Here, we follow the encoding of~\cite{simon-ijai01} to obtain the formulas of
varying size given the parameter $n$ of the encoder.
In the experiments, we generated 3 CNF formulas for each $n$ from 3 to 30
(i.e.\ $\text{\urq}_{n,i}$ for $n\in\{3,\ldots,30\}$ and $i\in\{1,2,3\}$),
which resulted in 84 instances.
As expected, the best performance on the URQ instances is demonstrated by zres.
Both \emph{maxhs} and \emph{lmhs} are not far behind.
Note that both \emph{maxhs} and \emph{lmhs} do exactly 1 call to CPLEX (due to
eq-seeding) after enumerating disjoint unsatisfiable cores.
This contrasts sharply with the poor performance of \emph{lp-wcnf}, which is
fed with the same problem instances.
Lingeling if augmented with Gaussian elimination (GA, see \emph{lgl} in
\autoref{fig:cactus-urq}) performs reasonably well being able to solve 29
instances.
However, as the result for \emph{lgl-nogauss} suggests, GA is crucial for
\emph{lgl} to efficiently decide URQ.
Note that \emph{lp-cnf} and \emph{cc-opb} are not shown in
\autoref{fig:cactus-urq} due to their inability to solve any instance.

The COMB benchmark set is supposed to inherit the complexity of both PHP and
URQ instances and contains formulas $\text{\php}_m^{m+1}\lor \text{\urq}_{n,i}$
with the PHP part being pairwise-encoded, where $m\in\{7,9,11,13\}$,
$n\in\{3,\ldots,10\}$, and $i\in\{1, 2, 3\}$, i.e.\ $|\text{COMB}|=96$.
As one can observe in \autoref{fig:cactus-comb}, even these small $m$ and $n$
result in instances that are hard for most of the competitors.
All IHS-based MaxSAT solvers (\emph{maxhs}, \emph{lmhs}, and \emph{lmhs-nes})
perform well and solve most of the instances.
Note that \emph{lgl} is confused by the structure of the formulas (neither CBR
nor GA helps it solve these instances). The same holds for \emph{zres}. As for
CPLEX, while \emph{lp-cnf} is still unable to solve any instance from the COMB
set, \emph{lp-wcnf} can also solve only 18 instances. The opposite observation
can be made for \emph{cc-cnf} and \emph{cc-opb}.

\subsubsection{Summary.}
As shown in \autoref{tab:summary}, given all the considered benchmarks sets,
the proposed problem transformation and the follow-up IHS-based MaxSAT solving
can cope with by far the largest number of instances overall (see the data for
\emph{maxhs}, \emph{lmhs}, and \emph{lmhs-nes}).
The core-guided and also resolution based MaxSAT solvers generally perform
well on the pigeonhole formulas (except \emph{wbo}, and this has to be
investigated further), which supports the theoretical claims of
papers.
However, using them does not help solving the URQ and also COMB benchmarks.
Also, as shown in \autoref{fig:php-nop}, the $\fml{P}$ clauses can be harmful
for MaxSAT solvers.
As expected, SAT solvers cannot deal with most of the considered formulas as
long as they do not utilize more powerful reasoning (e.g.\ GA or CBR).
However, and as the COMB instances demonstrate, it is easy to construct
instances that are hard for the state-of-the-art SAT solvers strengthened with
GA and CBR.
Finally, one should note the performance gap between \emph{maxhs} (also
\emph{lmhs}) and \emph{lp-wcnf} given that they solve the same instances by one
call to the same MIP solver with the only difference being the disjoint cores
precomputed by \emph{maxhs} and \emph{lmhs}.

\begin{table}[!t]
  \caption{Number of solved instances per solver.}
  \scriptsize
  \begin{adjustbox}{center}
   \setlength{\tabcolsep}{0.34em}
    \begin{tabular}{cccccccccccccccc}
      \toprule
      & &
      glucose &
      lgl &
      lgl-no\tablefootnote{This represents \emph{lgl-nogauss} for URQ and
      \emph{lgl-nocard} for PHP-pw, PHP-sc, and COMB.} &
      maxhs &
      lmhs &
      lmhs-nes &
      mscg &
      wbo &
      eva &
      lp-cnf &
      lp-wcnf &
      cc-cnf &
      cc-opb &
      zres \\
      \cmidrule{3-16}
      PHP-pw & {\tiny (46)} &
      7 &
      29 &
      7 &
      \textbf{46} &
      \textbf{46} &
      29 &
      \textbf{46} &
      10 &
      \textbf{46} &
      \textbf{46} &
      \textbf{46} &
      6 &
      5 &
      10 \\
      PHP-sc & {\tiny (46)} &
      13 &
      11 &
      11 &
      \textbf{46} &
      \textbf{46} &
      45 &
      \textbf{46} &
      15 &
      40 &
      \textbf{46} &
      \textbf{46} &
      6 &
      2 &
      8 \\
      URQ & {\tiny (84)} &
      3 &
      29 &
      4 &
      50 &
      44 &
      37 &
      5 &
      22 &
      3 &
      0 &
      6 &
      3 &
      0 &
      \textbf{84} \\
      COMB & {\tiny (96)} &
      11 &
      37 &
      41 &
      78 &
      \textbf{91} &
      80 &
      7 &
      13 &
      6 &
      0 &
      18 &
      6 &
      0 &
      39 \\
      \midrule
      Total & {\tiny (272)} &
      34 &
      106 &
      63 &
      220 &
      \textbf{227} &
      191 &
      104 &
      60 &
      95 &
      92 &
      116 &
      21 &
      7 &
      141 \\
      \bottomrule
    \end{tabular}
  \end{adjustbox} \label{tab:summary}
\end{table}

\jnoteF{Use no more than 4 pages in total:
  benchmarks+tables+plots+analysis.}


%
%
%

\section{Conclusions \& Research Directions} \label{sec:conc}

Resolution is at the core of CDCL SAT solving, but it also represents
its Achilles' heel.
Many crafted formulas are known to be hard for resolution, with
pigeonhole formulas representing a well-known
example~\cite{cook-jsl79}. More importantly, some of these examples
can occur naturally in some practical settings. 
In the context of \mxsat, researchers have proposed a dedicated form
of resolution, i.e.\ \mxsat
resolution~\cite{bonet-aij07,larrosa-aij08}, which was also shown not
to be more powerful than propositional resolution~\cite{bonet-aij07}
for the concrete case of pigeonhole formulas~\cite{cook-jsl79}.

This paper proposes a general transformation for CNF formulas, by
encoding the SAT decision problem as a \mxsat problem over Horn formulas. The 
transformation is based on the well-known dual-rail encoding, but it is
modified such that all clauses are Horn. More importantly, the paper
shows that, on this modified formula, \mxsat resolution can identify
in polynomial time a large enough number of empty soft clauses such
that this number implies the unsatisfiability of the original
pigeonhole formula. Furthermore, the paper shows that the same
argument can be used to prove a polynomial run time for the well-known
class of core-guided \mxsat solvers~\cite{mhlpms-cj13}.

Experimental results, obtained on formulas known to be hard for SAT
solvers, show that different families of \mxsat solvers perform far
better than the best performing SAT solvers, and also ILP solvers, on
these instances.

As the paper also hints at, future work will investigate effective
mechanisms for integrating Horn \mxsat problem transformation and
\mxsat reasoning techniques into SAT solvers. In contrast to cutting
planes or extended resolution, \mxsat algorithms already build on CDCL
SAT solvers; this is expected to facilitate integration.
Another research direction is to investigate similar transformations
for the many other examples for which resolution has exponential lower
bounds, but also when to opt to apply such transformations.

\jnoteF{Use no more than 0.5 pages in total.}



\bibliography{refs}
\bibliographystyle{abbrv}

\end{document}